\documentstyle[aps,prl,epsfig]{revtex}
\begin{document}
\title{Tenfold Magnetoconductance in a Non-Magnetic Metal Film}
\draft
\author{V.Yu. Butko\cite{Vladimir}, J.F. DiTusa, and P.W. Adams}
\address{Department of Physics and Astronomy\\Louisiana State University\\Baton Rouge, Louisiana,
70806}
\date{\today}
\maketitle
\begin{abstract}
We present magnetoconductance (MC) measurements of homogeneously
disordered Be films whose zero field sheet conductance (G) is described by the
Efros-Shklovskii hopping law $G(T)=(2e^2/h)\exp{-(T_o/T)^{1/2}}$. 
The low field MC of the films is negative with G decreasing $200\%$ below 1 T. 
In contrast the MC above 1 T is strongly positive.  At 8 T, G increases $1000\%$ in
perpendicular field and $500\%$ in parallel field.  In the simpler parallel case, we observe {\em
field enhanced} variable range hopping characterized by an attenuation of $T_o$ via the
Zeeman interaction.  
	\\     
\end{abstract}
\pacs{PACS numbers: 72.20.Ee, 71.30.+h, 73.50.-h}

	Over the last two decades a large and diverse research effort has grown around nanostructures
and low dimensional systems.  Not only has this area of investigation allowed the exploration of
fundamental quantum phenomena, but it also promises to impact magnetic storage technology and
magneto-electronics.  In particular, the development of magnetic superlattices has produced
encouraging increases in magnetic sensitivity by exploiting spin-dependent scattering of
electrons at the superlattice interfaces.  These systems can produce $500\%$
magnetoconductances in fields of a few telsa, and are commonly known as giant
magnetoresistance systems\cite{GMR1,GMR2}.  Extremely large magnetoresistance effects due to
magnetic scattering have also been observed in a variety of manganite systems \cite{CMR1,CMR2}. 
Interestingly though, there have been recent reports of large magnetoresistance effects in some
\emph{non-magnetic} systems.  The first is in the narrow band semiconductors
$Ag_{2+\delta}Se$ and $Ag_{2+\delta}Te$ which can have a
$200\%$ positive magnetoresistance at room temperature in fields of a few tesla
\cite{AgMR}.  The origin of the this effect is not understood. A second novel system is the 
newly discovered anomalous metallic phase in two-dimensional (2D) metal-oxide-semiconductor (MOSFET's)
devices \cite{Krav95}.  This metallic phase is observed to be extremely sensitive to magnetic
field and at low temperatures is suppressed by an arbitrarily weak field thereby producing a two
order of magnitude increase in resistance \cite{Sim97}.  There has been some conjecture that the
anomalous metallic phase, which is quite unexpected in a 2D system
\cite{LeeRam}, is stabilized by correlations
\cite{Dobro}.  If this is indeed the case, then the large magnetoresistances observed in the MOSFET
2D electron gases is, in fact, a many body effect.  This is intriguing in that it is a distinctly
different mechanism from the usual magnetic scattering processes that produce magnetoresistance
in magnetic films, and consequently it offers a novel strategy for realizing magnetic
sensitivity.  In the present Letter we describe an investigation of the field and temperature
dependent transport of highly disordered ultra-thin Be films.  Beryllium was chosen because one
can reproducibly grow smooth, non-granular films via thermal evaporation \cite{Be}.  This allowed
us to investigate field dependent electron correlation effects in a low atomic weight, high
carrier system, that is free from strong spin-orbit scattering, magnetic impurities, and grain
charging effects.  We have discovered that the hopping
transport in these films has an extreme sensitivity to magnetic field that can be attributed to
the convolution of $e-e$ correlations, disorder, and the carrier spin degrees of freedom.    

	 Beryllium forms smooth, dense, non-granular
films when thermally evaporated onto glass.  In fact, scanning force micrographs of the films'
exposed oxide surface did not reveal any salient morphological features down to our
resolution of 0.5 nm.  This non-granular morphology is crucial in that
it assures one that the measured resistance is representative of $e-e$
correlation effects and {\em not} grain charging effects. In extremely high resistance films any
significant granularity will result in field independent grain charging (Coulomb blockade)
effects preempting the many body effects of interest \cite{CB,EFT}.  This is also true of
electron tunneling measurements of the density of states (DOS) and indeed, we have recently made
the only direct measurement of the 2D Efros-Shklovskii Coulomb gap in Be films
\cite{CG}.  Another useful property of Be films is that they superconduct with a transition
temperature that is a monotonically decreasing function of the sheet resistance,
$T_c\approx0$ at $R\sim6$ k$\Omega$ \onlinecite{Be}.  By measuring the first-order
spin-paramagnetic parallel critical field transition in $\sim$1 k$\Omega$ films we were able to
demonstrate their extreme two-dimensionality and the absence of magnetic and spin-orbit scattering
\cite{Fulde,TM76}.  In the present study we used films with thicknesses ranging from
1.5 - 2.0 nm and corresponding low temperature sheet resistances ranging from
$R=10$k$\Omega$ to 2.6M$\Omega$.  They were deposited by thermally evaporating $99.5\%$ pure
beryllium metal onto fire polished glass substrates held at $84K$.  The evaporations were made in
a 4x$10^{-7}$ Torr vacuum at a rate $\sim0.15$ nm/s.  The film area was
1.5 mm x 4.5 mm. All of the samples discussed below where of sufficiently high resistance so
as to completely suppress the superconducting phase.  The film conductances were measured using
a standard four probe dc I-V technique.

	Electron correlations in strongly disordered electronic systems tend to produce a singular
depletion of the DOS near the Fermi surfaces. Efros and Shklovskii
\cite{ES1,ES2}  used a dimensionality argument to show that the Coulombic
interactions produce a linear Coulomb gap in the 2D DOS,
\begin{equation} 
N(eV)=\frac{\alpha(4\pi\epsilon_{o}\kappa)^2|eV|}{e^4},
\end{equation}
where $\kappa$ is the relative dielectric constant, $\epsilon_o$ is the permittivity
of free space, and
$\alpha$ is a constant of order unity.  The 3D Efros-Shklovskii Coulomb gap,
has a quadratic energy dependence and has recently been observed in Si:B \cite{MLee1,MLee2}. 
The gap described by Eq.(1), which we have demonstrated in these films
\onlinecite{CG}, produces a modified variable range hopping of the form,
\begin{equation} 
G(T)=G_o\exp{-(T_o/T)^{1/2}},
\end{equation}   
where G is the film sheet conductance, and $G_o$ is a constant. 
Deep in the hopping regime, $G_o$ is expected to be of the order of the quantum conductance $G_Q
=e^2/h$ \cite{ES3} and the correlation energy is given by
\begin{equation} 
T_o=2.8e^2/(k_B4\pi\epsilon_o\kappa\xi),
\end{equation} 
where $k_B$ is the Boltzman constant 
and $\xi$ is the localization length
\onlinecite{ES2}.  In the present Letter we present a systematic magnetoconductance
study of thin Be films in the hopping regime described by Eqs.(1)-(3).   
  
	Shown in the inset of Fig.\ 1 are the normalized magnetoconductances (MC) of a moderately
disordered 16k$\Omega$ (at $50mK$) film in both parallel and
perpendicular fields.  Note that the low field MC is negative but above ~1 T the MC becomes
positive for both field orientations.  Also note that even in this relatively low resistance film
the MC effects are of order $30\%$ of the total conductance.  This is to be compared with the
more typical $1\%$ MC magnitudes reported in metal films \cite{NLR89,FO88,Valles,Giordano}.  The
zero field G(T) of this film was stronger than the usual weak localization
$\ln(T)$ dependence \onlinecite{LeeRam} but was significantly weaker than ES behavior of Eq.(2).  
In higher resistance samples the MC field dependence was much stronger but
qualitatively similar.  For example, in the main body of Fig.\ 1 we show
the MC of a  3 M$\Omega$ film at 50mK.  In contrast to the
16k$\Omega$ film, this sample was well described by the ES hopping law of Eq.(2) with $T_o=1.6K$
and $G_o=2e^2/h$ (see Fig. 3). Both the ES behavior and the MC of Fig.\ 1 where quite reproducible
in films with $R > 1$ M$\Omega$.  

 There are several interesting features of the MC behavior that were observed in all of our
$R > 1$ M$\Omega$ samples.  The first was a factor of two decrease in the
conductivity as the field was increased from zero to ~0.5T.  The second was an astonishingly
large, positive, linear MC above 1T.  And finally, roughly a factor of two difference between
the high field parallel and perpendicular MC slopes. The solid lines in Fig.\ 1 are linear fits to
the high field data and have slopes 1/(1.1 T) and 1/(2.2 T) for the perpendicular and parallel
field data respectively. In the analysis given below we will primarily focus on the parallel
field MC behavior simply because it is the easiest to interpret. In parallel field the electron
trajectories do not accumulate Aharonov-Bohm phase and possibly subtle quantum interference
effects are avoided \cite{SS91}.  In essence parallel field affects the transport only through
the Zeeman energy of the electron spins.  With this simplification we outline a
phenomenological description of the MC behavior that only considers the energy levels of
localized states.  In this model localized states are characterized as being either unoccupied
(UO), singly occupied (SO), or doubly occupied (DO)
\cite{Kurobe}.  Due to the exclusion principle carriers can only hop to UO states or SO states of
opposite final spin, as we show schematically in Fig.\ 2.         

 We believe that the low field negative MC in Fig.\ 1 is a manifestation of the polarization
of correlated SO states by the applied field.  As the field polarizes the electron spins,
the density of opposite spin SO final states is reduced.  This in turn suppresses the SO to SO
hopping transitions \onlinecite{Kurobe}. If we assume that the SO states have quasi-free spins
then the suppression should be proportional to
\begin{equation}
P(H_{||},T)=\left[1-\tanh\left(\frac{\mu_BH_{||}}{k_B(T-\Theta)}\right)\right]
\end{equation} 
where $\mu_B$ is the Bohr magneton and $\Theta$ is the Weiss temperature.  If we
take $\Theta=0$ then $P(H_{||},T)$ is simply proportional to the free spin density oriented
counter to the field.  In Fig.\ 3 we have isolated the negative MC of Fig.\ 1 by subtracting off
the linear dependence of the parallel high field data. The solid line is a best single parameter
fit to Eq.(4) giving $\Theta=-93 mK$.  The negative Weiss temperature indicates a small
antiferromagnetic interaction among the localized states \cite{gfactor}. A similar fit to the
data in the inset of Fig.\ 1 gave
$\Theta=-50 mK$.  The quality of the fit in Fig.\ 3 is compelling evidence that the negative low
field MC is indeed due to SO polarization.  The fact that the conductance decreases by a
factor of two from zero field to 0.5T indicates that the SO to SO hopping events are a
significant part of the transport.

	We now turn our attention to the linear high field MC in Fig.\ 1.  We believe that this unusually
large positive MC may be associated with field ionization of DO sites.  If we assume that the
DO sites have a wide distribution of binding energies then there will be a significant number
of DO sites for which the first excited state is unbound.  Furthermore such weakly bound DO sites
are constrained to be in a spin-singlet state by the exclusion principle, therefore the
spin-triplet state is also unbound.  Consequently, when the Zeeman splitting is of order the
binding energy a DO site will ionize via the relaxation of the counter-aligned spin.  It is
natural to assume that the field ionization of DO sites will elevate carriers to UO sites at
higher energies thereby {\em increasing} the number of SO sites involved in hopping
conduction.  This should increase the localization length $\xi$ and possibly the dielectric
constant $\kappa$.  Therefore by Eq.(3) the overall correlation energy $T_o$ should decrease with
increasing field.  This is, in fact, the case as can be seen by the data in Fig.\ 4 where we have
made a semi-log plot of G as a function of $T^{-1/2}$ at several parallel fields.  The ES hopping
behavior of Eq.(2) is preserved in field as evidenced by the linearity of the 3T and 7T curves. 
The slopes of the curves are proportional to $T_o$ and clearly decrease with increasing field.
Thus the data in Fig.\ 4 can be characterized as field enhanced variable range hopping.

	Though the intercepts in Fig.\ 4 change slightly in field, we believe that the
parallel high field MC in Fig.\ 1 is almost completely dominated by field dependence of
$T_o(H_{||})$.  In fact, we can test the consistency of this conjecture by assuming that the high
field MC is linear in $H_{||}$, $G(H_{||},T)=G(0,T)\left[1+H_{||}/H_o\right]$ where $H_o$ is the
inverse of the slope in Fig.\ 1. We can now use this expression for $G(H_{||},T)$ to invert Eq.(2)
in order to obtain $T_o(H_{||})$,
\begin{equation}
T_{o}(H_{||})=\left[\sqrt{T_o(0)}+\sqrt{T}\ln\left(\frac{H_o}{H_o+H_{||}}\right)\right]^2.
\end{equation}         
In Fig.\ 5 we have plotted the values of $T_o$ extracted from curves such as those in Fig.\ 4 as a
function of parallel field.  The solid line is the field dependence of Eq.(5) with no adjustable
parameters.  The values of $H_o=2.2T$ and $T_o(0)=1.6K$ where obtained from the data in Figs.\ 1
and 4 respectively.  

	The outstanding agreement between Eq.(5) and the measured values of $T_o(H_{||})$
is strong evidence that the linear MC is indeed arising from the field dependence of
$T_o$.  Note that by Eq.(2) the magnitude of the MC could be much larger if it were
measured at lower temperatures.  Again, Eq.(5) was derived in order to demonstrate that the field
dependence of $T_o$ accounts for the high field MC in Fig.\ 1.  We do not know, however,
whether or not the MC will remain linear in $H_{||}$ at significantly higher sheet
resistances. 

	In conclusion, we find one of the largest MC's ever observed in a non-magnetic metal film.
The parallel field MC is a manifestation of the Zeeman splitting of correlated hopping channels.
We believe that the extreme uniformity of Be films unmasks this essential many body effect which
can be phenomenologically characterized by a field dependent $T_o$.  The extraordinary field
sensitivity of the hopping conductance highlights the dramatic role correlations and exchange
energies play in determining the localization and screening lengths in a highly disordered 2D
system.  Concurrent tunneling measurements of the DOS along with magnetotransport measurements of
$T_o$ should prove interesting and in principle would enable one to extract the field dependence
of the microscopic parameters of the theory such as
$\xi$ and $\kappa$.   
 
We gratefully acknowledge discussions with Boris Shklovskii, Vladimir Dobrosavljevic, and Boris
Altshuler.  We also thank Umit Alver for atomic force micrographs. This work was supported by the
NSF under Grant No.s DMR 99-72151 and DMR 97-02690.


%
\newpage
\begin{figure}
\vspace{0.5in}
\centerline{\epsfig{file=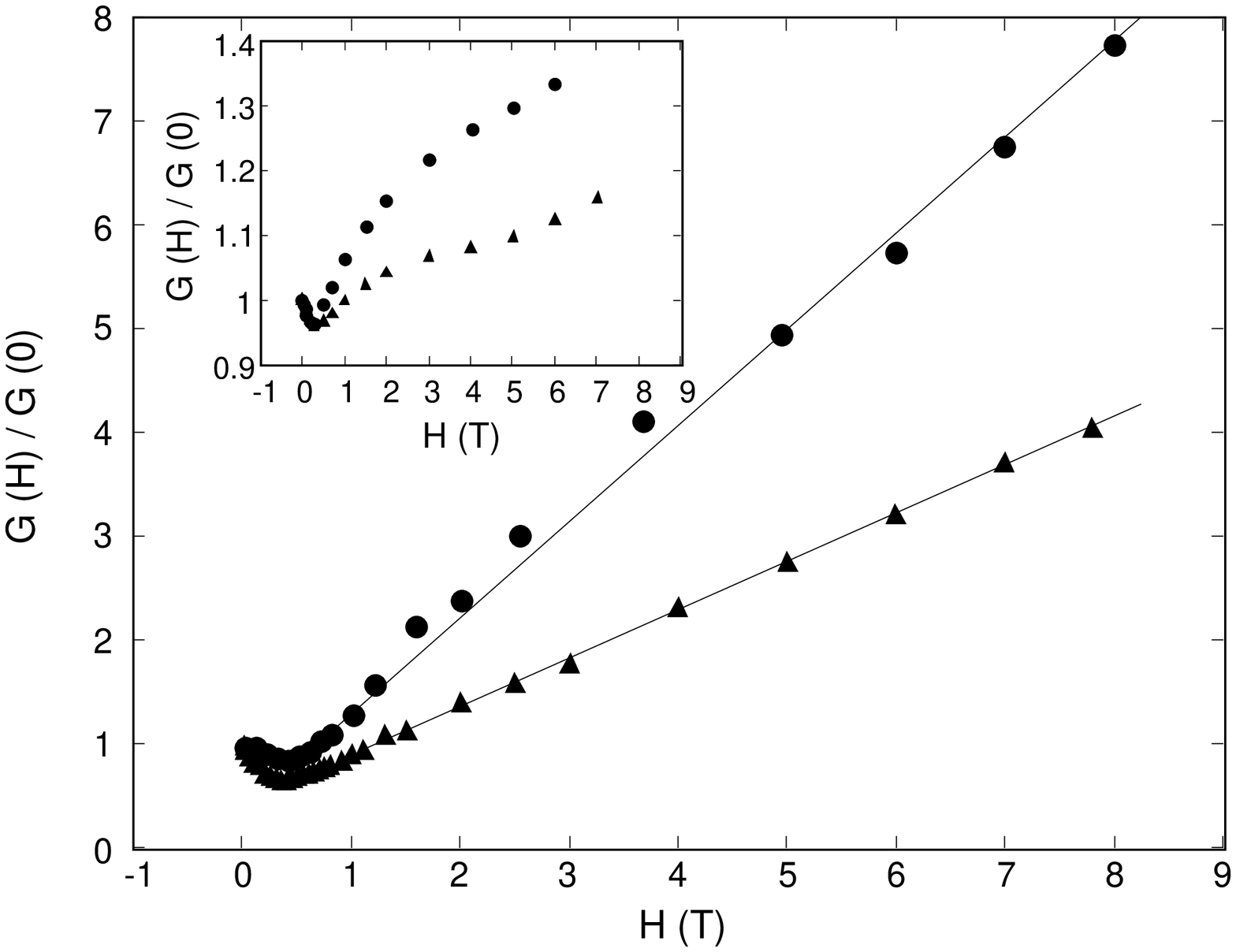,width=6.0in}}
\vspace{0.6in}
\caption{Relative magnetoconductance of a 3 M$\Omega$ Be film at 50mK. Circles: field
perpendicular to film surface. Triangles: field parallel to film surface. The solid lines are
linear fits to the data above $1 T$ with slopes of
$1/(1.1 T)$ and
$1/(2.2 T)$ for the perpendicular and parallel data respectively. Inset: relative
magnetoconductance of a 16k$\Omega$ Be film.}
 \label{Figure 1}
 \end{figure}
\newpage

\begin{figure}
\vspace{0.5in}
\centerline{\epsfig{file=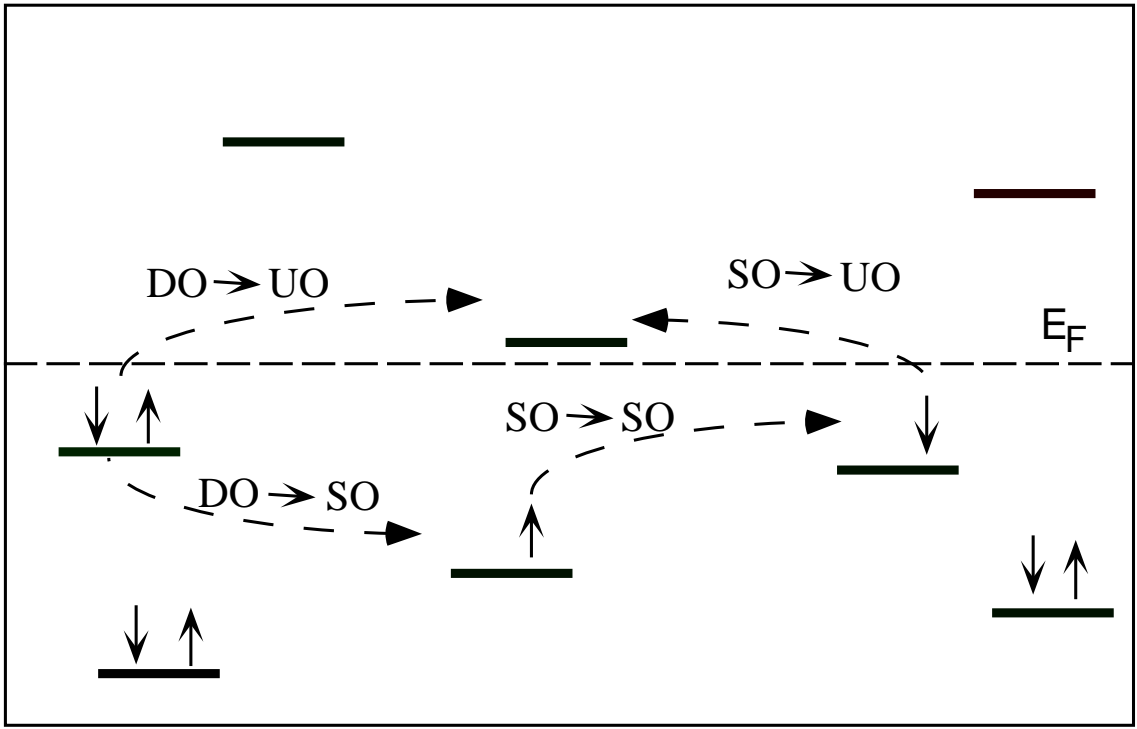,width=6.0in}}
\vspace{0.6in}
\caption{Schematic diagram showing the allowed hopping transitions between doubly occupied (DO),
singly occupied (SO), and unoccupied (UO) states near the Fermi energy.  We have neglected the
Coulomb interaction energy of DO sites.  An applied magnetic field will tend to polarize the
spins thereby cutting off the SO$\rightarrow$SO channel.  At the same time DO states can lower
their energy by ionizing (i.e., a DO$\rightarrow$UO transition with a spin flip).}
 \label{Figure 2}
 \end{figure}
\newpage

\begin{figure}
 \vspace{02.5in}
\centerline{\epsfig{file=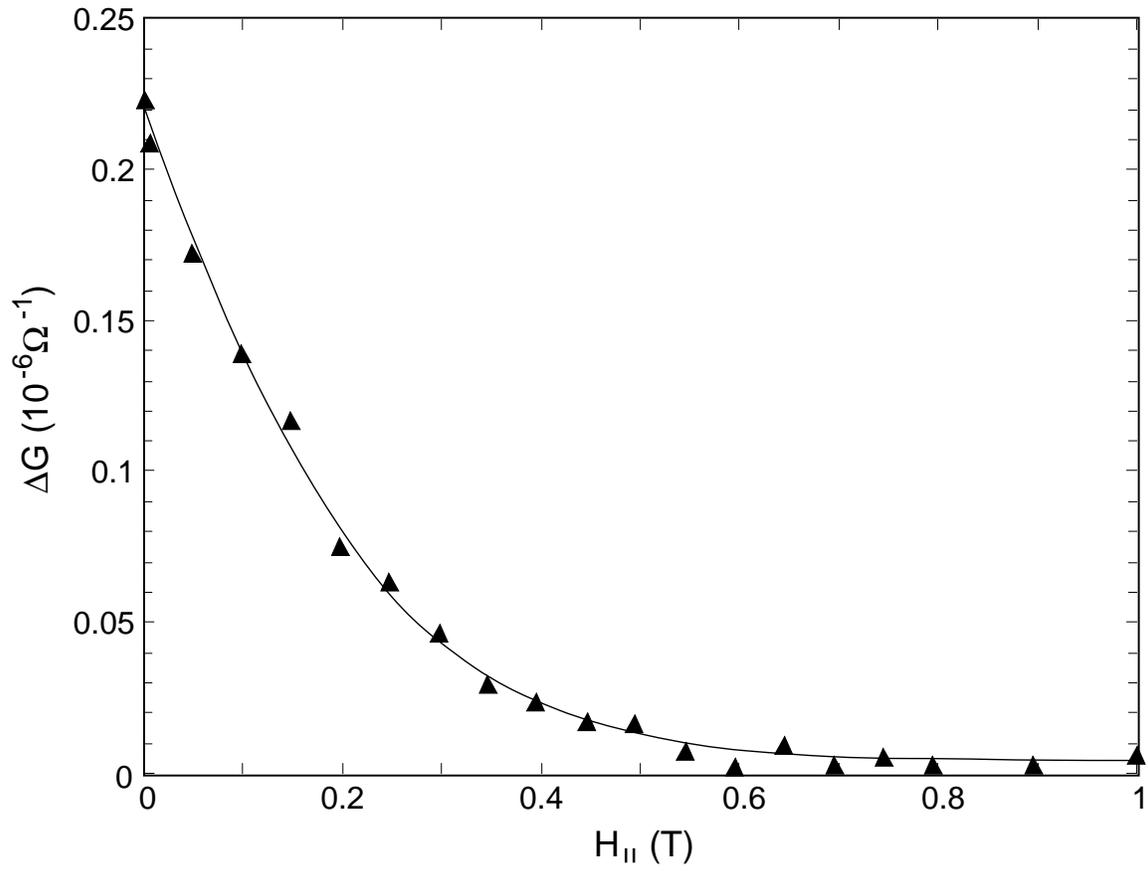,width=6.0in}}
\vspace{0.3in}
\caption{The parallel low field magetoconductance of the 3 M$\Omega$ film in Fig.\ 1 after
subtracting off the high field linear dependence.  The solid line is a fit to Eq.(4) were only
$\Theta$ was varied.}
 \label{Figure 3}
 \end{figure}
\newpage

\begin{figure}
 \vspace{2.5in}
\centerline{\epsfig{file=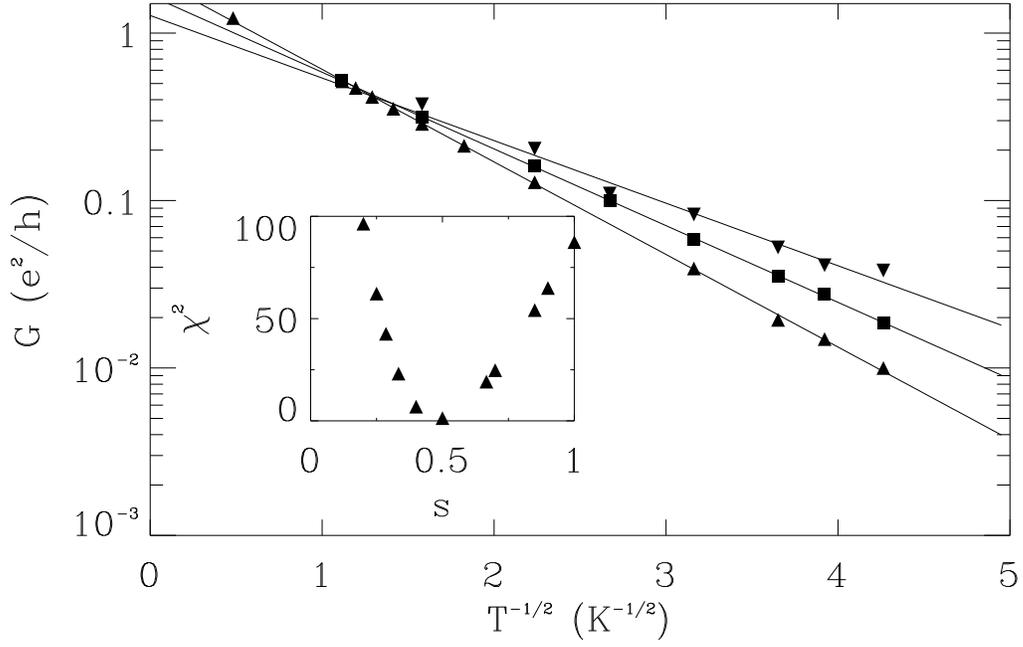,width=6.0in}}
\vspace{0.3in}
\caption{Semi-log plot of the film conductance as function of $T^{-1/2}$ at three different
parallel magnetic fields. Up triangles: $H_{||}=0$. Squares: $H_{||}=3.0 T$. Down triangles:
$H_{||}=7.0 T$.  The solid lines are linear fits to the data from which $T_o(H_{||})$ were
obtained via Eq.(2).}
 \label{Figure 4}
 \end{figure}
\newpage

\begin{figure}
\vspace{2.5in}
\centerline{\epsfig{file=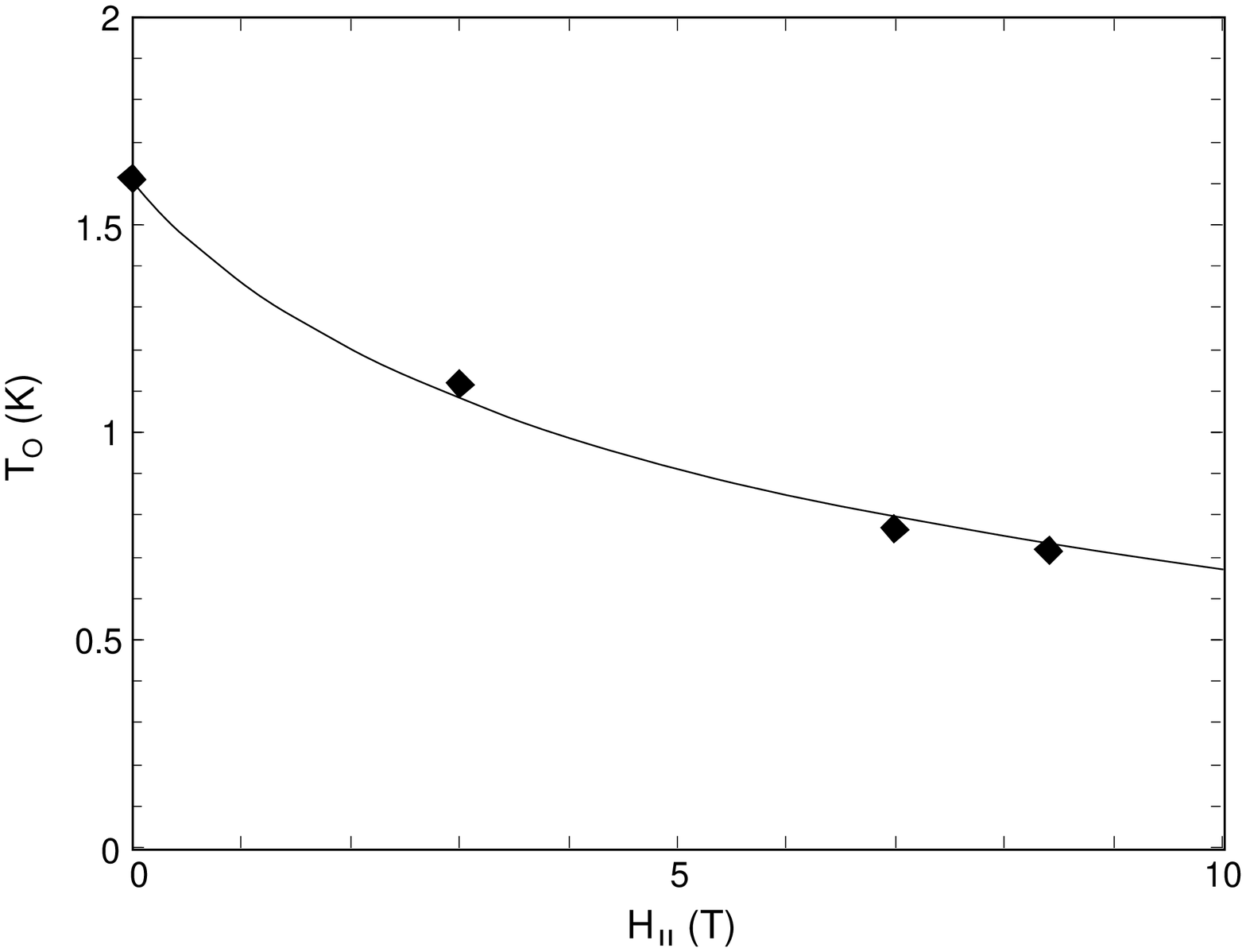,width=6.0in}}
\vspace{0.3in}
\caption{Measured values of $T_o$ as a function of $H_{||}$.  The solid line is the prediction of
Eq.(5) with no adjustable parameters.}
 \label{Figure 5}
 \end{figure}
%

%


\begin{references}

\bibitem[*] {Vladimir} Permanent address: Ioffe Physical Technical Institute (PTI), Russian
Academy of Sciences, Polytekhnicheskaya Street, 26, 194021, St. Petersburg, Russia.
 
\bibitem{GMR1} M.N. Baibich {\it et al.}, Phys. Rev. Lett. {\bf61}, 2472 (1988).
\bibitem{GMR2} R.E. Camley and R.L. Stamps, J. Phys. Condens. Mat. {\bf5}, 3727 (1993).
\bibitem{CMR1} S. Jin {\it et al.}, Science {\bf264}, 413 (1994).
\bibitem{CMR2} Y. Tokura {\it et al.}, J. Appl. Phys. {\bf79}, 5288 (1996).
\bibitem{AgMR} R. Xu {\it et al.}, Nature {\bf390}, 57 (1997).
\bibitem{Krav95} S.V. Kravchenko, W.E. Mason, G.E. Bowker, J.E. Furneaux, V.M.
Pudalov, and M. D'Iorio, Phys. Rev. B {\bf51}, 7038 (1995).
\bibitem{Sim97} D. Simonian, S.V. Kravachenko, M.P. Sarachik, and V.M. Pudalov,
Phys. Rev. Lett. {\bf79}, 2304 (1997).
\bibitem{LeeRam} P.A. Lee and T.V. Ramakrishnan, Rev. Mod. Phys. {\bf57}, 287
(1985).
\bibitem{Dobro} V. Dobrosavljevic, E. Abrahams, E. Miranda, and S. Chakravarty,
Phys. Rev. Lett. {\bf79}, 455 (1997).
\bibitem{Be} P.W. Adams, P. Herron, and E.I. Meletis, Phys. Rev. B {\bf58}, 2952
(1998).
\bibitem{CB} D.V. Averin and K.K. Likharev, in {\it Mesoscopic Phenomena in Solids}, edited by B.
Altshuler, P. Lee, and R. Webb (Elsevier, Amsterdam, 1991), Chap. 6.
\bibitem{EFT} Wenhao Wu and P.W. Adams, Phys. Rev. B {\bf50}, 13065 (1994).
\bibitem{CG} V.Yu. Butko, J.F. Ditusa, and P.W. Adams, submitted.
\bibitem{Fulde} P. Fulde, Adv. Phys. {\bf22}, 667 (1973).
\bibitem{TM76} P.M. Tedrow and R. Meservey, Phys. Lett. {\bf 58A}, 237
(1976).
\bibitem{ES1} A.L. Efros and B.I. Shklovskii, J. Phys. C {\bf8}, L49 (1975).
\bibitem{ES2} B.I. Shklovskii and A.L. Efros, {\it Electronic Properties of Doped
Semiconductors}, (Springer, New York, 1984).
\bibitem{MLee1} J.G. Massey and Mark Lee, Phys. Rev. Lett. {\bf75}, 4266 (1995).
\bibitem{MLee2} J.G. Massey and Mark Lee, Phys. Rev. Lett. {\bf77}, 3399 (1996).
\bibitem{ES3} R.Berkovits and B.I.Shklovskii, J. Phys. Cond. Mat. {\bf11}, 779
(1998).
\bibitem{NLR89} M. Nissim, Y. Lereah, and R. Rosenbaum, Phys. Rev. B {\bf40}, 6351 (1989).
\bibitem{FO88} O. Faran and Z. Ovadyahu, Phys. Rev. B {\bf38}, 5457 (1988).
\bibitem{Valles} Shih-Ying Hsu and J.M. Valles, Jr., Phys.
Rev. Lett. {\bf74}, 2331 (1995).
\bibitem{Giordano} N. Giordano and M.A Pennington, Phys. Rev. B {\bf47}, 9693 (1993).
\bibitem{SS91} B.I. Shklovskii and B.Z. Spivak, in {\it Hopping Transport in Solids}, edited by M.
Pollak and B.I. Shklovskii (Elsievier Science Publishers B.V., New York 1991), p.271.


\bibitem{Kurobe} A. Kurobe, J. Phys. C.: Solid State Phys. {\bf19}, 2201 (1986); M. Eto, Phys.
Rev. B {\bf48}, 4933 (1993).
\bibitem{gfactor} $\Theta < 0$ is also consistent with recent evidence that the Land\'{e} g-factor
is less than 2 in Al and Be films, see V.Yu. Butko, P.W. Adams, and I.L. Aleiner, Phys. Rev.
Lett. {\bf82}, 4284 (1999); P.W. Adams and V.Yu. Butko, {\it Proceedings of the 22nd International
Conference on Low Temperature Physics}, Physica B, in press. 










\end{references}
\end{document}